\documentclass[iop]{emulateapj}

\usepackage{apjfonts}
\usepackage{graphicx}
\usepackage{epstopdf}
\usepackage{ulem}
\usepackage[utf8]{inputenc}
\usepackage[T1]{fontenc}
\usepackage{url}
\usepackage{hyperref}
\usepackage{mdwlist}
\usepackage{amsmath} 
\usepackage{enumitem}

\makeatletter
\def\url@leostyle{%
 \@ifundefined{selectfont}{\def\UrlFont{\sf}}{\def\UrlFont{\small\ttfamily}}}
\makeatother
\begin{document}

\newcommand{\ls}{{_<\atop^{\sim}}}
\newcommand{\gs}{{_>\atop^{\sim}}}
\def \spose#1{\hbox  to 0pt{#1\hss}}  
\def \ls{\mathrel{\spose{\lower 3pt\hbox{$\sim$}}\raise  2.0pt\hbox{$<$}}}
\def \gs{\mathrel{\spose{\lower  3pt\hbox{$\sim$}}\raise 2.0pt\hbox{$>$}}}
\newcommand{\Ha}{\hbox{{\rm H}$\alpha$}}
\newcommand{\Hb}{\hbox{{\rm H}$\beta$}}
\newcommand{\OIII}{\hbox{[{\rm O}\kern 0.1em{\sc iii}]}}
\newcommand{\NII}{\hbox{[{\rm N}\kern 0.1em{\sc ii}]}}
\newcommand{\angstrom}{\textup{\AA}}
\newcommand\ionn[2]{#1$\;${\scshape{#2}}}

\font\btt=rm-lmtk10


\title{Stellar Populations of Barred Quiescent Galaxies}

\shorttitle{Stellar Populations of Barred Quiescent Galaxies}

\shortauthors{Cheung et al.}


\author{Edmond Cheung\altaffilmark{1} \altaffilmark{2} \dag, Charlie Conroy\altaffilmark{3}, 
E. Athanassoula\altaffilmark{4}, 
Eric F. Bell\altaffilmark{5}, 
A. Bosma\altaffilmark{4},
Carolin N. Cardamone\altaffilmark{6},
S. M. Faber\altaffilmark{1} \altaffilmark{7},
David C. Koo\altaffilmark{1} \altaffilmark{7},
Chris Lintott\altaffilmark{8} \altaffilmark{9}, 
Karen L. Masters\altaffilmark{10} \altaffilmark{11},
Thomas Melvin\altaffilmark{10},
Brooke Simmons\altaffilmark{8},
Kyle W. Willett\altaffilmark{12}
} 

\altaffiltext{1}{Kavli Institute for the Physics and Mathematics of the Universe (WPI), The University of Tokyo Institutes for Advanced Study, The University of Tokyo, Kashiwa, Chiba 277-8583, Japan}
\altaffiltext{2}{Department of Astronomy and Astrophysics, 1156 High Street, University of California, Santa Cruz, CA 95064}
\altaffiltext{3}{Harvard-Smithsonian Center for Astrophysics, 60 Garden St., Cambridge,
MA, USA}
\altaffiltext{4}{Aix Marseille Universit\'e, CNRS, LAM (Laboratoire d'Astrophysique de Marseille) UMR 7326, 13388, Marseille, France}
\altaffiltext{5}{Department of Astronomy, University of Michigan, 500 Church St., Ann Arbor, MI 48109, USA}
\altaffiltext{6}{Department of Science, Wheelock College, Boston, MA 02215, USA}
\altaffiltext{7}{UCO/Lick Observatory, Department of Astronomy and Astrophysics, University of California, 1156 High Street, Santa Cruz, CA 95064}
\altaffiltext{8}{Oxford Astrophysics, Department of Physics, University of Oxford, Denys Wilkinson Building, Keble Road, Oxford OX1 3RH}
\altaffiltext{9}{Astronomy Department, Adler Planetarium and Astronomy Museum, 1300 Lake Shore Drive, Chicago, IL 60605, USA}
\altaffiltext{10}{Institute of Cosmology \& Gravitation, University of Portsmouth, Dennis Sciama Building, Portsmouth, PO1 3FX, UK}
\altaffiltext{11}{SEPnet, South East Physics Network}
\altaffiltext{12}{Minnesota Institute for Astrophysics, School of Physics and Astronomy, University of Minnesota, MN 55455, USA}
\altaffiltext{\dag}{ec2250@gmail.com}

\slugcomment{Accepted for publication by ApJ}


\begin{abstract}

Selecting centrally quiescent galaxies from the Sloan Digital Sky Survey (SDSS) to create high signal-to-noise ($\gs100$ \AA$^{-1}$)  stacked spectra with minimal emission line contamination, we accurately and precisely model the central stellar populations of barred and unbarred quiescent disk galaxies. By splitting our sample by redshift, we can use the fixed size of the SDSS fiber to model the stellar populations at different radii within galaxies. At $0.02<z<0.04$, the SDSS fiber radius corresponds to $\approx1$ kpc, which is the typical half-light radii of both classical bulges and disky pseudobulges. Assuming that the SDSS fiber primarily covers the bulges at these redshifts, our analysis shows that there are no significant differences in the stellar populations, i.e., stellar age, [Fe/H], [Mg/Fe], and [N/Fe], of the {\it bulges} of barred vs. unbarred quiescent disk galaxies. Modeling the stellar populations at different redshift intervals from $z=0.020$ to $z=0.085$ at fixed stellar masses produces an estimate of the stellar population {\it gradients} out to about half the typical effective radius of our sample, assuming null evolution over this $\approx1$ Gyr epoch. We find that there are no noticeable differences in the slopes of the azimuthally averaged gradients of barred vs. unbarred quiescent disk galaxies. These results suggest that bars are not a strong influence on the chemical evolution of quiescent disk galaxies.

 \end{abstract}



\section{Introduction} \label{sec:introduction}

\setcounter{footnote}{12}

The stellar populations of galaxies offer a unique view of the formation and evolution of galaxies. In this work, we use stellar populations to explore the evolution of barred quiescent disk galaxies. 

Bars are predicted to affect the properties of their host galaxies through the slow rearrangement of energy and mass, a process called secular evolution \citep{kk04, athanassoula13b, sellwood14}. This redistribution of material leads to many effects, including the flattening of the initial global abundance gradients of their host galaxies \citep{friedli94, friedli95, martinet97, minchev10, dimatteo13, kubryk13}, which results in bulge abundances that are dependent on the presence of star formation. Specifically, if star formation is present, then the gaseous metallicities of bulges in barred galaxies are predicted to increase compared to unbarred galaxies \citep{friedli94, friedli95, martel13}, while the bulge stellar metallicities stay relatively unchanged \citep{friedli94}. In this case, bars are also predicted to produce younger stellar populations at the center and the ends of the bar structure \citep{wozniak07}. But if star formation is absent, then both the gaseous and stellar metallicities of bulges in barred galaxies are predicted to decrease due to dilution by lower metallicity stars and gas from further galactocentric distances, assuming an initial negative metallicity gradient \citep{friedli94, dimatteo13}. 

So far, works that have studied the {\it bulge gas-phase metallicities} of barred galaxies have only considered star-forming galaxies, which means that the bulge gaseous metallicities of barred galaxies are expected to be enhanced relative to unbarred galaxies. While there are works that agree with this prediction, there are others that do not. For example, \cite{ellison11}, who used a large sample of galaxies from the Sloan Digital Sky Survey (SDSS), found that the gas-phase metallicities in the inner few kpc of barred galaxies are higher (by $\sim$0.06 dex) than unbarred galaxies at a given stellar mass. But \cite{cacho14}, who also used a large sample of SDSS galaxies, found that for a given stellar mass there are no significant differences in the gas-phase metallicities at the inner few kpc of barred and unbarred galaxies \citep[see also][]{henry99}. However, in contrast to all these results, \cite{dutil99} and \cite{considere00} found that the central gas-phase metallicities of barred galaxies are lower than that of unbarred galaxies. 

Works that have studied the {\it bulge stellar populations} of barred galaxies have considered both star-forming and quiescent galaxies. However, in analyzing their results, most of these works did not separate star-forming galaxies from quiescent galaxies, thus producing results that are difficult to interpret. Nonetheless, we summarize their findings for the benefit of the reader: \cite{moorthy06} found that barred galaxies have higher bulge stellar metallicities than unbarred galaxies at a given velocity dispersion, which is consistent with the findings of \cite{perez11}, who additionally found that the bulges of a sample of barred S0 galaxies are slightly $\alpha$-enhanced compared to that of unbarred S0 galaxies. These results, however, are in contrast to those of \cite{coelho11}, \cite{williams12}, and \cite{cacho14}; they all found no difference in the stellar metallicity of bulges. Moreover, there is a similar disagreement about the bulge stellar ages of barred and unbarred galaxies: \citealt{perez11}, \citealt{williams12}, and  \citealt{cacho14} found no difference, while \cite{coelho11} did. 

Clearly, there is no consensus on whether bars affect the bulge abundances of their host galaxies. However, the lack of any strong bulge abundance differences in these works suggests that any potential effects from bars are probably small.  

\begin{figure*}[t!] 
\centering
\includegraphics[scale=.62]{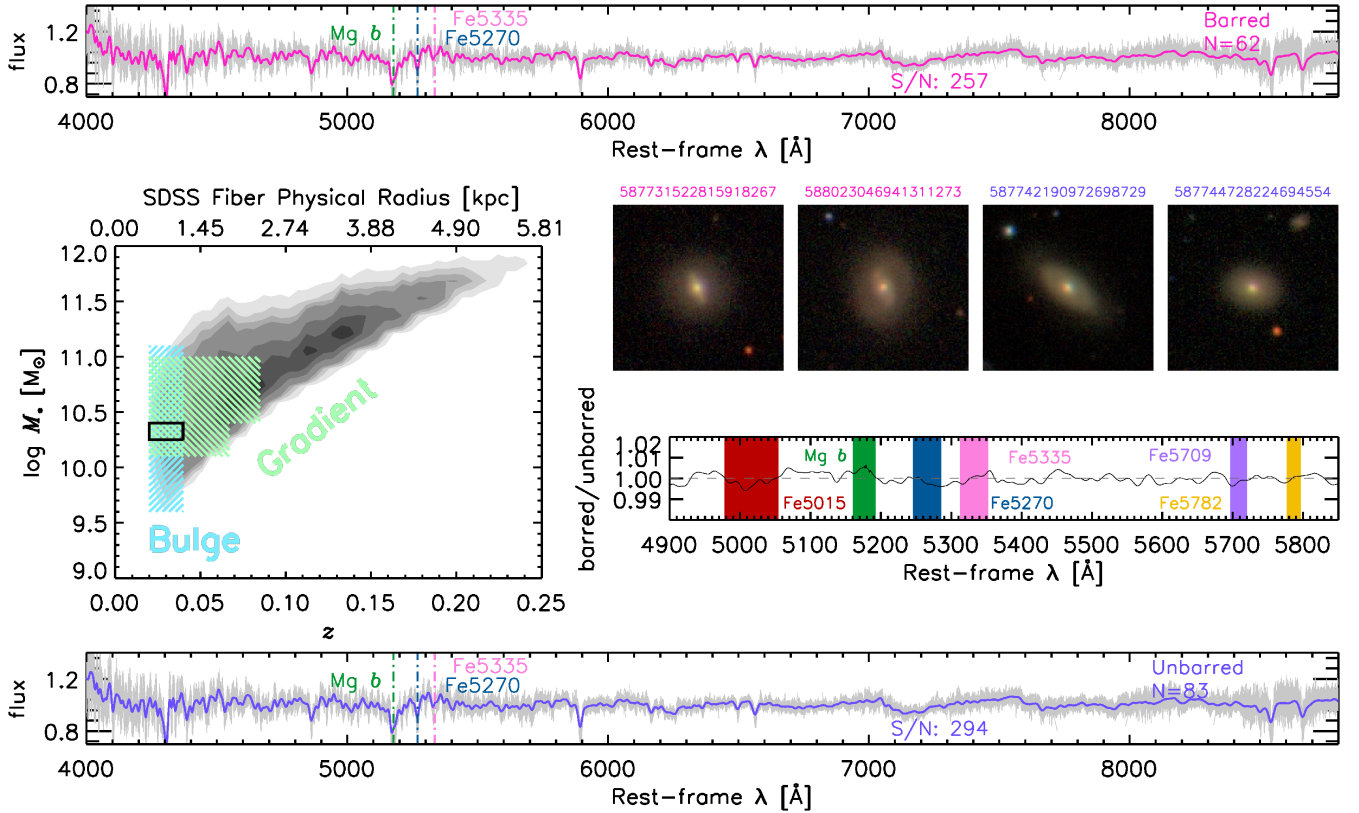}
\caption{Graphic summary of this study. Stellar mass ($M_*$) vs. redshift ($z$) of our quiescent galaxy sample is shown in the left-center panel. Light-blue hatches mark the galaxies used to analyze the bulge stellar populations, and the light-green hatches mark the galaxies used for the stellar population gradient analysis. The spectra of the barred and unbarred galaxies in the black rectangle at $0.02<z<0.04$ and at $10.25<\log~M_*/M_{\odot}<10.40$ are shown at the top and bottom, respectively. Each individual spectrum is plotted in grey, and the stacked spectra of the barred and unbarred galaxies are plotted in their respective colors. These stacks have the high signal-to-noise (S/N $\geq100$ \AA$^{-1}$) needed to obtain accurate and precise stellar population parameters. Example images of barred and unbarred galaxies from the highlighted bin are also displayed, along with their DR7 SDSS object IDs. Finally, the center-right panel shows the ratio of the barred and unbarred stacked spectra at several Lick indices. This spectral comparison illustrates the extremely subtle differences that are analyzed in this work, and hence the need for high S/N and full-spectrum modeling. 
\label{fig:massvsz}}
\end{figure*} 

Turning to global gradients, which are predicted to be flatter in barred galaxies regardless of the presence of star formation, reveals a similarly confused landscape. While some studies found that the {\it gas-phase metallicity gradients} of barred galaxies are flatter than unbarred galaxies \citep{vila-costas92, martinroy94, zaritsky94}, other studies found no difference \citep{sanchez14, ho15}. 

Studies on the {\it stellar population gradients} of barred galaxies have also found conflicting results. Using slits placed along the bars of 20 galaxies, \cite{perez09} found a variety of stellar metallicity and age gradients, including positive, negative, and null gradients \citep[see also][]{perez07}. For two of these galaxies, \cite{sanchez-blazquez11} found that the stellar metallicity and age gradients are flatter along the bar than along the disk.  In agreement, \cite{williams12} used a sample of 28 galaxies to show that the stellar metallicity gradients of boxy/peanut-shaped bulges (which are presumably, indicative of the presence of a bar; \citealt{athanassoula05}) are shallower than that of a sample of unbarred early-type galaxies. But most recently, using 62 face-on spiral galaxies from the CALIFA survey \citep{sanchez12}, \cite{sanchez-blazquez14} found no difference in the stellar metallicity gradient or age gradient (their gradients extend out to the disk region) between barred and unbarred galaxies.

Similar to bulge abundances, there is no consensus on whether bars affect the abundance gradients of their host galaxies, indicating that any potential effects due to bars are small. Therefore highly accurate and precise chemical analysis is needed to robustly detect them. 

This work aims to achieve this goal by using very high signal-to-noise (S/N $\geq100$ \AA$^{-1}$) stacks of barred and unbarred galaxies, identified by Galaxy Zoo 2 \citep{willett13}, to model their stellar populations. Aside from the unprecedented S/N stacks of barred galaxies, an additional important and unique feature of our study is the minimization of emission lines by selecting only quiescent galaxies. This latter feature circumvents the need to model the emission line contribution, which is complex and introduces more uncertainty to the resulting stellar population parameters \citep{conroy13}. Moreover, with quiescent galaxies, we are comparing to predictions that have not been investigated, i.e., simulations without star formation.

We describe the data, sample selection, stacking, and stellar population synthesis model in \S2 and \S3. \S4 presents our results, which indicates that there are no significant differences in the stellar populations of barred and unbarred quiescent disk galaxies. We present a comparison to past works in \S5 and discuss our results in \S6. Finally, we close in \S7. Throughout this paper, we assume a flat cosmological model with $H_{0} = 70$ km s$^{-1}$ Mpc$^{-1}$, $\Omega_{m} = 0.30$, and  $\Omega_{\Lambda} =0.70$, and all magnitudes are given in the AB magnitude system. 

\section{Data} \label{sec:data}

Since our study aims to model the stellar populations of barred and unbarred quiescent disk galaxies, we must procure spectra of a large sample of these types of galaxies. Thus we use the spectroscopic Main Galaxy Sample in the Legacy area of the SDSS Data Release Seven (SDSS DR7; \citealt{strauss02, abazajian09}); these spectra are measured with a 3$\arcsec$ fiber, hence the results presented in this work only concern the central regions of our galaxies. 

\subsection{Sample selection} \label{sub:sample_selection}

In order to most accurately and most precisely model the detailed abundance patterns of galaxies, passive galaxies with little-to-no emission lines are required. Thus we select quiescent galaxies as described by \cite{peek10}, i.e., no H$\alpha$ or [\ionn{O}{ii}]$\lambda$3727 emission; these criteria have been shown to exclude star-forming galaxies and AGN hosts \citep{graves09}. Compared to a red sequence defined by specific star formation rate or rest-frame $g-r$, this spectroscopic quiescent criteria is much more strict---about $50\%$ of the galaxies in the red sequence make it into our quiescent sample. However, the fraction of face-on disk galaxies and the fraction of barred galaxies, which are defined below, are similar in both the red sequence and our spectroscopically-defined quiescent sample, indicating that our sample selection is not picking a special subset of red sequence galaxies. The presence of quiescent disks is consistent with recent work that showed that galaxies with disks have made up a significant fraction of the red sequence since $z\sim1$ \citep{bundy10, masters10}. 

In addition to our quiescent selection, we follow the SDSS DR7 caveats\footnote{\href{http://classic.sdss.org/dr7/products/spectra/index.html\#caveats}{http://classic.sdss.org/dr7/products/spectra/index.html\#caveats}} and select galaxies with median S/N>10 \AA$^{-1}$ and with velocity dispersions of $70 ~{\rm km~s^{-1}} < \sigma < 420~ {\rm km~s^{-1}}$. For these selections, we have utilized the MPA-JHU DR7 value-added catalogs\footnote{\href{http://www.mpa-garching.mpg.de/SDSS/DR7/}{http://www.mpa-garching.mpg.de/SDSS/DR7/}}. 
 
\subsubsection{Selecting barred and unbarred disk galaxies} \label{subsub:bar_selection}

To select barred disk galaxies from our quiescent sample, we first select face-on disk galaxies based on the criteria of \cite{masters11} and \cite{cheung13}. Namely, we select galaxies with axis ratios ($b/a$) larger than 0.5 since it is difficult to identify bars in edge-on galaxies ($b/a$ measurements are from the GIM2D single S\'ersic fits by \citealt{simard11}). We then use the debiased\footnote{The debiasing procedure accounts for the deterioration of the image quality due to increasing distance of galaxies, i.e., with increasing redshift. The underlying assumption is that galaxies of a similar luminosity and size will share the same average mix of morphologies within the epoch covered by our redshift range ($0.020<z<0.085$; see \citealt{willett13} for more details).} Galaxy Zoo 2 (GZ2) visual morphological classifications from \cite{willett13} to select non-edge-on disk galaxies by requiring that for each galaxy, at least a quarter of its classifications involved answering the bar question. 

A relevant aside is to compare our quiescent disk sample to the red spiral sample of \cite{masters10}: we find that less than $1\%$ of their sample make it into our sample. The main reason is our requirement of little-to-no H$\alpha$ or [\ionn{O}{ii}]$\lambda$3727 emission, which is apparently present in almost all red spiral galaxies (\cite{masters10} also excluded large bulges and imposed a more strict inclination cut). This comparison indicates that the disks of this sample of galaxies are distinctive in that they have little to no detectable spiral structure, i.e., they are featureless disks.

From this face-on quiescent disk sample, we select barred galaxies with a debiased bar vote fraction threshold of 0.5, i.e., galaxies with $p_{\rm bar}>0.5$ are considered barred. This bar threshold has been shown to be a reliable indicator of strong bar features---almost all galaxies with $p_{\rm bar}\ge0.5$ were classified as possessing a strong bar by \cite{nair10b} (see \citealt{masters12} and \citealt{willett13})---and has been adopted by several past Galaxy Zoo works \citep{masters11, masters12,  melvin14, cheung15}. Using a more strict debiased bar vote fraction threshold of $p_{\rm bar}=0.8$ does not change our qualitative results. We present representative images of barred galaxies in a given stellar mass and redshift interval in Fig.~\ref{fig:gradients_with_stacked_bars} to illustrate the reliability of the GZ2 visual classifications.

To select unbarred disk galaxies, we choose galaxies with a debiased bar vote fraction of zero, i.e., $p_{\rm bar}=0.0$, from the face-on quiescent disk sample described above. This bar vote fraction choice is different from previous Galaxy Zoo works because we want to minimize contamination from weak bars, which has been shown by \cite{masters12} and \cite{willett13} to likely have $0.1<p_{\rm bar}<0.5$. 

The selection of face-on unbarred quiescent disks is difficult and there is no perfect technique. However, comparing the structural parameters (global S\'ersic index, central surface stellar mass density, central velocity dispersion, and bulge-to-total ratio) of our barred and unbarred quiescent disk samples for a given stellar mass bin shows a similar distribution, indicating that we select similar types of galaxies (see Figs. \ref{fig:faceondisk_vs_elliptical_0} and \ref{fig:faceondisk_vs_elliptical_1} in Appendix \ref{appen:cheng_selection}). Moreover, comparing the structural parameters of our quiescent disk samples to a sample of quiescent spheroid-dominated galaxies shows differences at the lowest stellar masses, indicating that our samples are not severely contaminated by spheroid-dominated objects. 

And finally, since our study concerns isolated galaxies, we eliminate merging galaxies by discarding all galaxies with a Galaxy Zoo merging parameter, $p_{\rm mg}$, larger than 0.4 \citep{darg10}. 

\begin{deluxetable*}{ccccc}
\tabletypesize{\scriptsize}
\tablewidth{0pt} 
\tablecaption{Properties of Stacked Spectra}
\tablehead{   
\colhead{} &
\colhead{$z$ Bin} &
\colhead{$M_*$ Bin} &
\colhead{Number\tablenotemark{a}} &
\colhead{S/N\tablenotemark{b}}\\
\colhead{} &
\colhead{} &
\colhead{($M_{\odot}$)} &
\colhead{} &
\colhead{({\AA}$^{-1}$)} 
} 
\startdata 
Bulge-barred 1 & $0.020<z<0.040$ & $9.60 < \log~M_* < 9.90$ & 13 & 80 \\
Bulge-barred 2 & $0.020<z<0.040$ & $9.90 < \log~M_* < 10.10$ & 32 & 130 \\
Bulge-barred 3 & $0.020<z<0.040$ & $10.10 < \log~M_* < 10.25$ & 39 & 170 \\
Bulge-barred 4 & $0.020<z<0.040$ & $10.25 < \log~M_* < 10.40$ & 62 & 257 \\
Bulge-barred 5 & $0.020<z<0.040$ & $10.40 < \log~M_* < 10.55$ & 71 & 307 \\
Bulge-barred 6 & $0.020<z<0.040$ & $10.55 < \log~M_* < 10.70$ & 66 & 307 \\
Bulge-barred 7 & $0.020<z<0.040$ & $10.70 < \log~M_* < 10.85$ & 26 & 189 \\
Bulge-barred 8 & $0.020<z<0.040$ & $10.85 < \log~M_* < 11.10$ & 10 & 142 \\ \noalign{\smallskip} \hline \noalign{\smallskip} 
Bulge-unbarred 1 & $0.020<z<0.040$ & $9.60 < \log~M_* < 9.90$ & 39 & 133 \\
Bulge-unbarred 2 & $0.020<z<0.040$ & $9.90 < \log~M_* < 10.10$ & 81 & 216 \\
Bulge-unbarred 3 & $0.020<z<0.040$ & $10.10 < \log~M_* < 10.25$ & 84 & 253 \\
Bulge-unbarred 4 & $0.020<z<0.040$ & $10.25 < \log~M_* < 10.40$ & 83 & 294 \\
Bulge-unbarred 5 & $0.020<z<0.040$ & $10.40 < \log~M_* < 10.55$ & 87 & 317 \\
Bulge-unbarred 6 & $0.020<z<0.040$ & $10.55 < \log~M_* < 10.70$ & 40 & 248 \\
Bulge-unbarred 7 & $0.020<z<0.040$ & $10.70 < \log~M_* < 10.85$ & 18 & 178 \\
Bulge-unbarred 8 & $0.020<z<0.040$ & $10.85 < \log~M_* < 11.10$ & 10 & 141 \\ \\ \noalign{\smallskip} \hline \hline\noalign{\smallskip} \\

Gradient-barred 1 & $0.020<z<0.040$ & $10.10 < \log~M_* < 10.40$ & 101 & 306 \\
Gradient-barred 2 & $0.040<z<0.055$ & $10.10 < \log~M_* < 10.40$ & 131 & 272 \\
Gradient-barred 3 & $0.055<z<0.067$ & $10.10 < \log~M_* < 10.40$ & 59 & 153 \\ \\
Gradient-barred 4 & $0.020<z<0.040$ & $10.40 < \log~M_* < 10.70$ & 137 & 435 \\
Gradient-barred 5 & $0.040<z<0.055$ & $10.40 < \log~M_* < 10.70$ & 139 & 338 \\
Gradient-barred 6 & $0.055<z<0.067$ & $10.40 < \log~M_* < 10.70$ & 143 & 261 \\
Gradient-barred 7 & $0.067<z<0.077$ & $10.40 < \log~M_* < 10.70$ & 148 & 235 \\
Gradient-barred 8 & $0.077<z<0.085$ & $10.40 < \log~M_* < 10.70$ & 66 & 158 \\ \\
Gradient-barred 9 & $0.020<z<0.040$ & $10.70 < \log~M_* < 11.00$ & 36 & 233 \\
Gradient-barred 10 & $0.040<z<0.055$ & $10.70 < \log~M_* < 11.00$ & 90 & 318 \\
Gradient-barred 11 & $0.055<z<0.067$ & $10.70 < \log~M_* < 11.00$ & 92 & 262 \\
Gradient-barred 12 & $0.067<z<0.077$ & $10.70 < \log~M_* < 11.00$ & 86 & 213 \\
Gradient-barred 13 & $0.077<z<0.085$ & $10.70 < \log~M_* < 11.00$ & 94 & 207 \\   \noalign{\smallskip} \hline \noalign{\smallskip}
Gradient-unbarred 1 & $0.020<z<0.040$ & $10.10 < \log~M_* < 10.40$ & 167 & 383 \\
Gradient-unbarred 2 & $0.040<z<0.055$ & $10.10 < \log~M_* < 10.40$ & 509 & 527 \\
Gradient-unbarred 3 & $0.055<z<0.067$ & $10.10 < \log~M_* < 10.40$ & 341 & 373 \\ \\
Gradient-unbarred 4 & $0.020<z<0.040$ & $10.40 < \log~M_* < 10.70$ & 127 & 399 \\
Gradient-unbarred 5 & $0.040<z<0.055$ & $10.40 < \log~M_* < 10.70$ & 429 & 604 \\
Gradient-unbarred 6 & $0.055<z<0.067$ & $10.40 < \log~M_* < 10.70$ & 687 & 623 \\
Gradient-unbarred 7 & $0.067<z<0.077$ & $10.40 < \log~M_* < 10.70$ & 842 & 600 \\
Gradient-unbarred 8 & $0.077<z<0.085$ & $10.40 < \log~M_* < 10.70$ & 441 & 411 \\ \\
Gradient-unbarred 9 & $0.020<z<0.040$ & $10.70 < \log~M_* < 11.00$ & 26 & 213 \\
Gradient-unbarred 10 & $0.040<z<0.055$ & $10.70 < \log~M_* < 11.00$ & 152 & 418 \\
Gradient-unbarred 11 & $0.055<z<0.067$ & $10.70 < \log~M_* < 11.00$ & 292 & 493 \\
Gradient-unbarred 12 & $0.067<z<0.077$ & $10.70 < \log~M_* < 11.00$ & 423 & 509 \\
Gradient-unbarred 13 & $0.077<z<0.085$ & $10.70 < \log~M_* < 11.00$ & 472 & 484 
\enddata
\tablenotetext{a}{Total number of galaxies in the stacked spectrum.}
\tablenotetext{b}{Effective median S/N of the stacked spectrum.}
\label{tab:stacks}
\end{deluxetable*}

\subsubsection{Quiescent barred galaxies} \label{sub:quiescent_selection}
 
Although barred galaxies are commonly thought of as star-forming, recent works have shown that barred galaxies are actually more frequent among the red sequence disks than the blue cloud \citep{masters11, lee12, cheung13}. And when considering absolute numbers in the volume-limited sample of \cite{cheung13}, the numbers of red and blue barred galaxies are comparable. Finally, although our spectroscopically-defined quiescent sample is approximately half of the photometrically-defined red sequence, the bar fraction of the two samples is similar, indicating that our quiescent barred galaxy sample is not an unusual class of red sequence barred galaxies.  

\subsection{Stacking} 
Because detailed abundance analysis requires high S/N spectrum (S/N $\geq100$ \AA$^{-1}$; \citealt{cardiel98}), we need to stack many SDSS spectra (each SDSS spectra has a typical S/N$\approx20$ \AA$^{-1}$) to achieve the necessary S/N \citep[e.g.,][]{graves09}. 

We use two stacking schemes that correspond to our two science goals: (1) bins of stellar mass at $0.02<z<0.04$ for our bulge analysis and (2) bins of redshift at three mass intervals for our gradient analysis (see Table \ref{tab:stacks}; galaxy stellar masses [not fiber stellar masses] are taken from the MPA-JHU DR7 value-added catalogs); we explain the motivations for these schemes in \S\ref{sec:results}. 

Each individual spectrum was continuum-normalized using gaussian convolution, smoothed to an effective velocity dispersion of 300 km s$^{-1}$, corrected for galactic extinction using the parameterization of \cite{cardielli89}, and flux-calibrated according to \cite{yan11}. The empirical flux calibration from \cite{yan11} results in relative spectrophotometric precision of 0.1\%, increasing the accuracy and precision of our model fitting, which we describe in the next section. Each spectrum contributes equally to a stack, but portions of spectra that are 5$\sigma$ deviations from the initial stack are trimmed. Only a small percentage of spectra are trimmed in this manner, and the exclusion of this trimming does not affect our final results. The typical S/N of our stacks is $\gs100$ \AA$^{-1}$; see Table \ref{tab:stacks} for each stack's S/N. 

We compare the stacks of our barred and unbarred samples in Fig.~\ref{fig:stack_comparison_0} and \ref{fig:stack_comparison_1} of Appendix \ref{appen:stacks}. Focusing on the ratio of the stacks at the Lick indices, we find that there are no significant differences, which is supported by Fig. \ref{fig:lick_indices}, which plots several Lick indices as a function of mass for barred and unbarred quiescent disk galaxies. This initial analysis supports our findings using full-spectrum fitting in \S\ref{sec:results}. 

The total number of barred and unbarred galaxies that make up the stacks for our bulge analysis is 319 and 442, respectively. The total number of barred and unbarred galaxies that make up the stacks for our gradient analysis is 1,322 and 4,908, respectively.

\section{Model Fitting} \label{sec:model_fitting}
We fit our high S/N stacked spectra with the latest version of the stellar population synthesis (SPS) model of \cite{conroy12}, which is detailed in \cite{conroy14} \citep[see also][]{choi14}. We follow the methodology of \cite{conroy14} in this present work.

Briefly, this SPS model uses three sets of stellar isochrones and the MILES \citep{sanchez-blazquez06} and IRTF \citep{cushing05, rayner09} empirical stellar spectral libraries to fit for the full inputted spectrum at four wavelength intervals: 4000 \AA--4700 \AA, 4700 \AA--5700 \AA, 5700 \AA--6400 \AA, and 8000 \AA--8800 \AA. The wavelength range 6400 \AA--8000 \AA~is not included because it is at the edges of the MILES and IRTF wavelength coverage. A two-part Kroupa IMF is assumed. Non-solar abundance patterns are modeled with response functions using the ATLAS12/SYNTHE code suites \citep{kurucz70, kurucz93} and applied differentially onto the template model galaxy spectrum. A Markov Chain Monte Carlo (MCMC) algorithm was used to explore the 40 free parameters of the model, which include: redshift, velocity dispersion, two population ages (the age of the dominant population and the age of the younger population)\footnote{Although a quiescent disk galaxy may have a more complex star formation history, the old ages we derive in \S\ref{sec:results} indicates that this issue should not affect our results \citep{serra07}.}, the mass fraction of the younger population, four nuisance parameters\footnote{These parameters are meant to describe uncertain aspects of SPS modeling. Two of these are the temperature and the fractional flux from young stars or hot horizontal branch stars. One allows for the addition of arbitrary amounts of M giant light, and the final one is a shift in effective temperature (see \citealt{conroy12b} and \citealt{conroy14} for more details).}, 13 emission line strengths, the velocity broadening of the emission lines, and the abundances of C, N, Na, Mg, Si, Ca, Ti, V, Cr, Mn, Fe, Co, Ni, Sr, and Ba, and O, Ne, S (the latter three are varied in lock-step). The systematic uncertainties (e.g., incomplete stellar spectral libraries, unknown aspects of stellar evolution, metallicity evolution) are probably $\sim 0.05$ dex \citep{conroy14, choi14}.

We illustrate the quality of our model fits in Figs.~\ref{fig:spectra_comparison_0} and \ref{fig:spectra_comparison_1} of Appendix \ref{appen:residuals}; it shows that the fits are excellent and moreover, the quality of the barred and unbarred fits are indistinguishable. 

\section{Results} \label{sec:results}

\begin{figure*}[t!] 
\centering
\includegraphics[scale=.62]{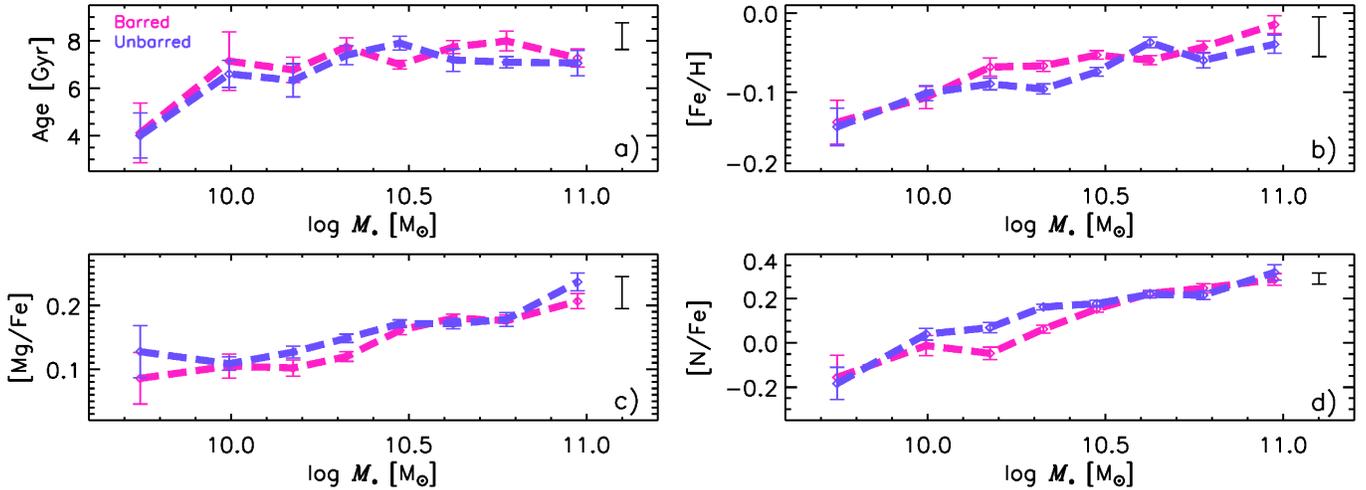}
\caption{Bulge stellar populations vs. stellar mass for barred and unbarred quiescent disk galaxies. Namely, {\it a):} Age vs. $\log~M_*$, {\it b):} [Fe/H] vs. $\log~M_*$, {\it c):} [Mg/Fe] vs. $\log~M_*$, and {\it d):} [N/Fe] vs. $\log~M_*$ at $0.02<z<0.04$. The error estimates are based on the full posterior distributions of each parameter from the MCMC spectrum-fitting algorithm. We display a 0.05 dex floating error bar in each panel as a conservative estimate of the systematic errors, which should be assumed in addition to the displayed error bars. There are no significant differences in age, [Fe/H], [Mg/Fe], or [N/Fe] of the bulges of barred and unbarred quiescent disk galaxies. 
\label{fig:stellarpop_vs_mass}}
\end{figure*}

\begin{figure*}[t!] 
\centering
\includegraphics[scale=.6]{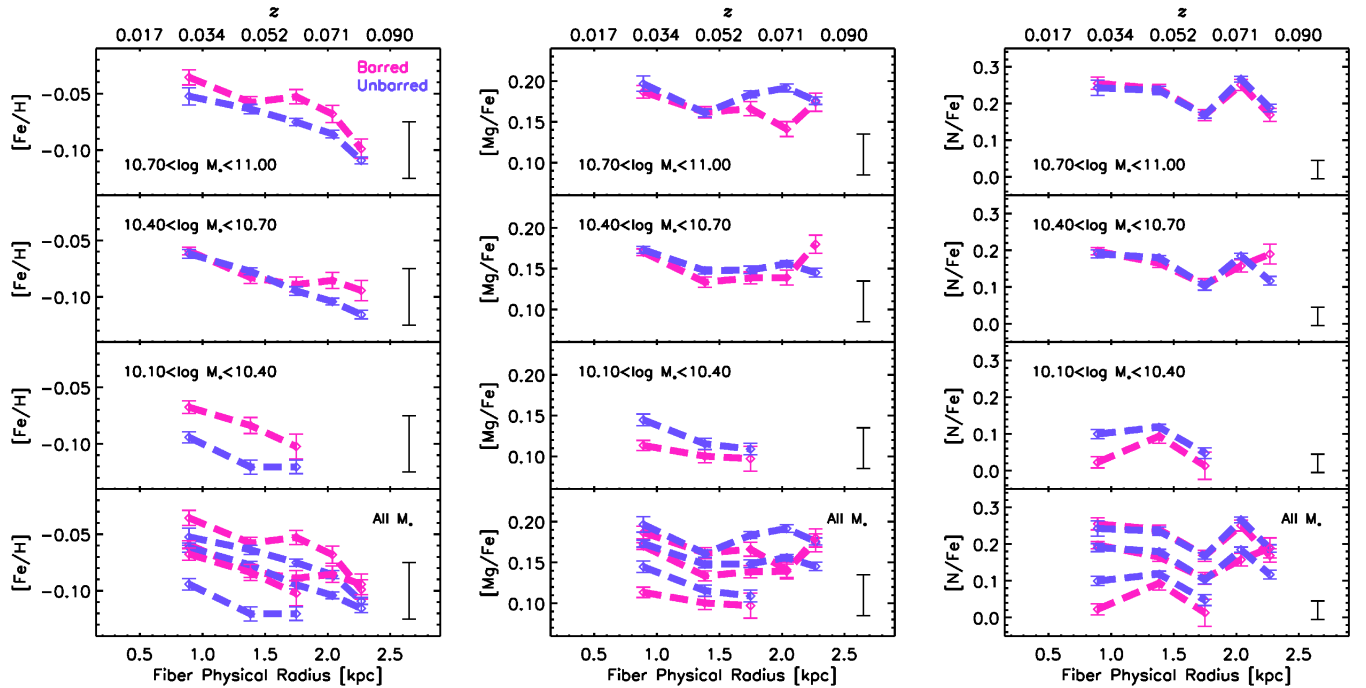}
\caption{Stellar population gradients of [Fe/H], [Mg/Fe], and [N/Fe] at three different stellar mass bins---$10.1<\log~M_*/M_{\odot}<10.4$, $10.4<\log~M_*/M_{\odot}<10.7$, and $10.7<\log~M_*/M_{\odot}<11.0$---for barred and unbarred quiescent disk galaxies. The fourth row superimposes all mass bins. The corresponding redshifts are displayed on the top axis. We display a 0.05 dex floating error bar in every panel as a conservative estimate of the systematic errors. There are no clear differences in the slopes of gradients of barred and unbarred galaxies. 
\label{fig:gradients}}
\end{figure*} 

\subsection{Stellar populations of the bulges of barred and unbarred galaxies} \label{sub:bulge_pop}

To model the stellar populations of galaxy bulges, our analysis only considers redshifts between $0.02<z<0.04$ since the SDSS fiber radius ($1\farcs5$) at $z=0.04$ corresponds to $\approx 1$ kpc, which is the typical half-light radii of both classical bulges and disky pseudobulges \citep{fisher10}. The lower limit of $z=0.02$ corresponds to the redshift where the SDSS spectrograph starts to cover [\ionn{O}{ii}]$\lambda$3727. The sample for our bulge stellar population analysis is highlighted by the blue hatches in the $\log~M_*-z$ distribution of our quiescent sample in Fig.~\ref{fig:massvsz}. The stellar masses we consider range from $\log ~M_*/M_{\odot}=9.6$ to $\log ~M_*/M_{\odot}=11.1$, which we split into eight $\log~M_*$ bins; these bins are listed in the top rows of Table \ref{tab:stacks} and are labeled ``Bulge-barred'' or ``Bulge-unbarred''. The barred and unbarred stacked spectra of the highlighted bin ($0.02<z<0.04$ \& $10.25<\log~M_*/M_{\odot}<10.40$) are shown in the upper and lower panels of Fig.~\ref{fig:massvsz}, with each spectrum plotted in grey. Two images of barred and unbarred galaxies from the highlighted bin are also shown. The ratio of the barred and unbarred stacked spectra at several Lick indices are highlighted in the center-right panel. There are extremely subtle differences in these Lick indices, indicating that: (1) high S/N is needed to detect these differences and (2) full-spectrum modeling can help bring out features that are not covered by Lick indices. Appendix \ref{appen:stacks} presents a more detailed comparison of the stacks and their residuals---they show similar features to the comparison in Fig.~\ref{fig:massvsz}.  

With these high S/N stacks, we model the bulge stellar populations of our barred and unbarred quiescent disk galaxies at each stellar mass bin. Our results are shown in Fig.~\ref{fig:stellarpop_vs_mass}; it displays the stellar age, [Fe/H] (a metallicity indicator), [Mg/Fe] (a star formation timescale indicator), and [N/Fe] (another star formation timescale indicator) vs. stellar mass of barred quiescent disks in magenta and unbarred quiescent disks in purple (this is the color scheme throughout the paper). There is a general correlation between all stellar population parameters and stellar mass for both populations, similar to the trends between stellar population parameters and central velocity dispersion of early-type galaxies \citep[e.g.,][]{thomas05}.

The errors estimates shown in Fig.~\ref{fig:stellarpop_vs_mass} are based on the full posterior distributions of each fitted parameter from the MCMC spectrum-fitting algorithm. Systematic uncertainties are $\sim 0.05$ dex \citep{conroy14, choi14} and should be assumed in addition to the displayed error estimates. Thus the main result of Fig.~\ref{fig:stellarpop_vs_mass} is that there are no significant differences in the stellar populations of the bulges of barred vs. unbarred quiescent disk galaxies, which is supported by the $P$-values of $\ge 0.14$ (based on the Z-score).

\subsection{Stellar population gradients } \label{sub:gradient}

Since the SDSS fiber size corresponds to a larger physical size at higher redshift, we can probe stellar populations at larger physical galactocentric radii by analyzing higher redshift galaxies \citep[e.g.,][]{yan12}. Combining this approach with the assumption that galaxy properties (e.g., galaxy size, bar length, and their abundance patterns) do not evolve significantly at a fixed stellar mass over our redshift range of $0.020<z<0.085$, which corresponds to $\approx 1$ Gyr, we can estimate stellar population gradients. Our upper limit of $z=0.085$ corresponds to the redshift limit of the debiasing procedure applied to the GZ2 vote fractions \citep{willett13}. Therefore, we separate our sample into seven bins of redshift, from $z=0.020$ to $z=0.085$, at three stellar mass bins:  $10.1<\log~M_*/M_{\odot}<10.4$, $10.4<\log~M_*/M_{\odot}<10.7$, and $10.7<\log~M_*/M_{\odot}<11.0$; these bins are listed in the bottom rows of Table \ref{tab:stacks} and are labeled ``Gradient-barred'' or ``Gradient-unbarred.'' This sample is highlighted with green hatches in the $\log~M_*$ vs. $z$ panel of Fig.~\ref{fig:massvsz}. 

 \begin{figure*}[t!] 
\centering
\includegraphics[width=\textwidth]{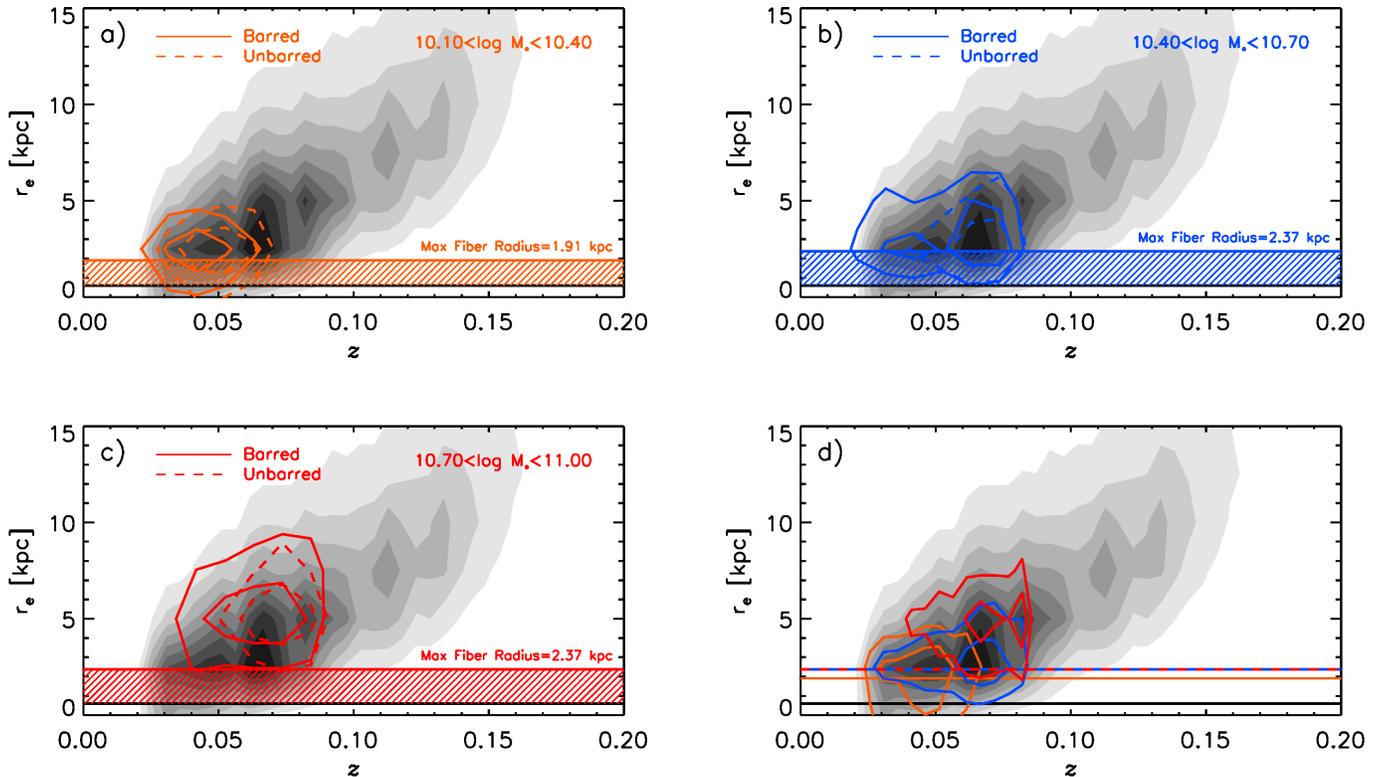}
\caption{The semi-major half-light radius ($r_{\rm e}$) vs. redshift ($z$) distribution of our quiescent sample is displayed in grey contours in each panel. Panels a-c overplots the $r_{\rm e}-z$ distribution of the three mass bins of barred and unbarred galaxies used in the stellar population gradient analysis The radius range that the SDSS fiber can cover for a given mass bin is highlighted by the colored hatches in panels a-c. Panel d overplots the $r_{\rm e}-z$ distribution of the three mass bins (combining both barred and unbarred galaxies) and their lines of maximum SDSS fiber physical radius in their respective colors; the line of the minimum SDSS fiber physical radius is shown as the black horizontal line. The maximum SDSS fiber physical radius is $\approx50-70\%$ of the median $r_{\rm e}$ of each mass bin.   
\label{fig:re_vs_z}}
\end{figure*} 

\subsubsection{SDSS fiber size vs. typical galaxy size} 

Before we present the stellar population gradients, an important question that needs to be addressed is ``how much of our galaxies is covered by the SDSS fiber?'' The answer is shown in Fig.~\ref{fig:re_vs_z}, which plots semi-major half-light radius ($r_{\rm e}$; from the GIM2D single S\'ersic fits by \citealt{simard11}) vs. redshift ($z$). Each panel displays the  $r_{\rm e}-z$ distribution of our quiescent sample (see \S\ref{sub:sample_selection}) in the grey contours. Overplotted in panels a-c are the $r_{\rm e}-z$ distributions of the three mass bins considered in the stellar population gradient analysis. The barred and unbarred samples are displayed in solid and dashed contours, respectively; there is no evident difference in their $r_{\rm e}-z$ distribution. Moreover, there is no clear $r_{\rm e}$ evolution at a given mass bin, supporting our assertion of null $r_{\rm e}$ evolution in our redshift range. 

Within panels a-c, the colored hatched lines represent the range that is covered by the SDSS fiber at a given mass bin, which, as shown in Fig.~\ref{fig:massvsz}, corresponds to different maximum redshifts. Panel d overplots all the $r_{\rm e}-z$ distributions of the three mass bins (combining both barred and unbarred galaxies) and their lines of maximum SDSS fiber physical radius in their respective colors. 

Fig.~\ref{fig:re_vs_z} shows that for each mass bin, the maximum SDSS fiber physical radius is $\approx50-70\%$ of the median $r_{\rm e}$, meaning that the gradients in Fig.~\ref{fig:gradients} extend out to $\approx50-70$\% the median galaxy semi-major half-light radius. Thus even at the highest redshift bin, our gradients are still dominated by the central regions of our galaxies, indicating that our gradient analysis is not sensitive to the outskirts of our galaxies.

\subsubsection{SDSS fiber size vs. typical bar length}

\begin{figure*}[t!] 
\centering
\includegraphics[scale=1]{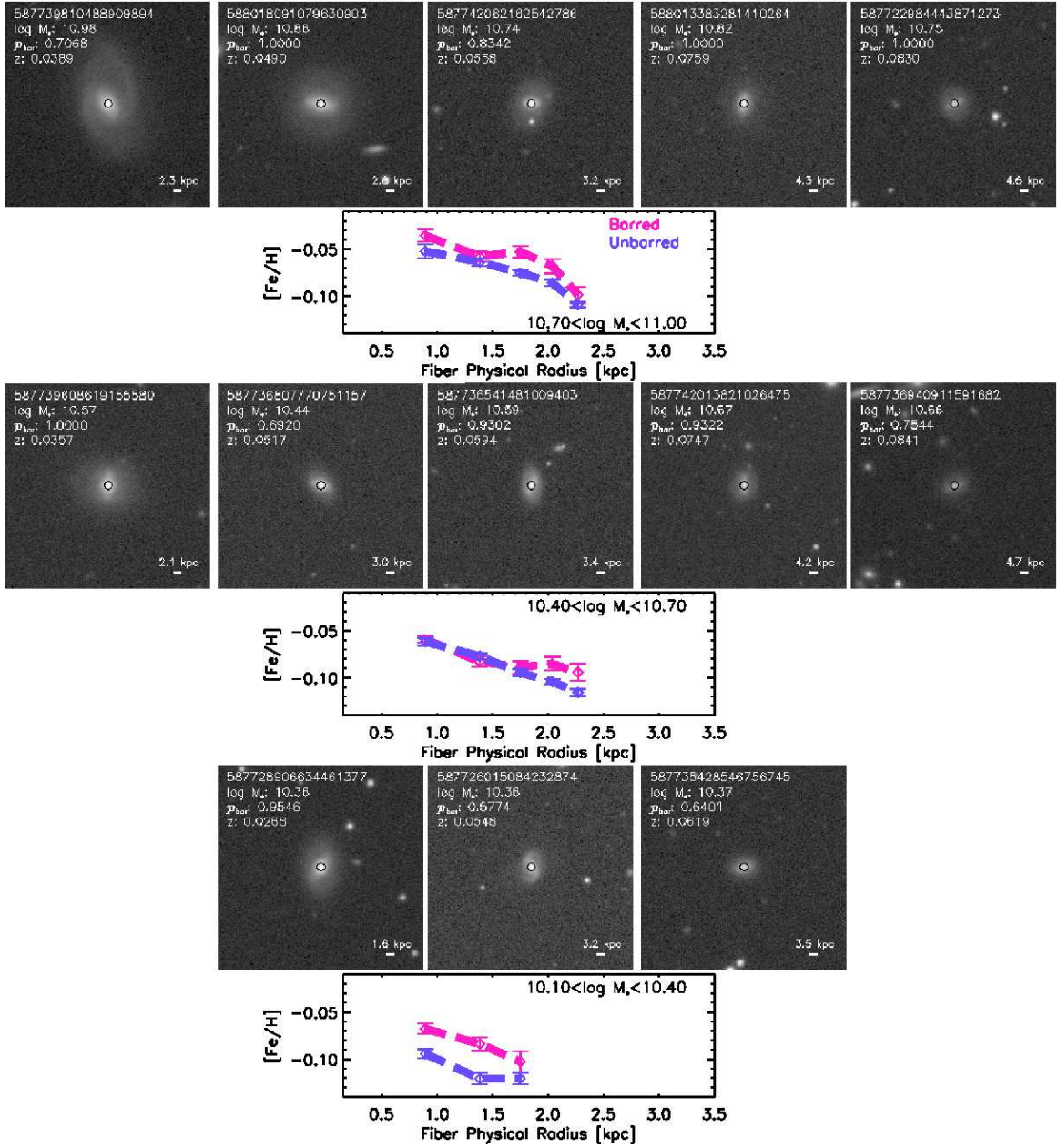}
\caption{An illustration of the relative size of the SDSS fiber to the bar structure as a function of redshift. SDSS $r$-band images of typical barred galaxies at each redshift interval for a given stellar mass are shown at the top of their corresponding [Fe/H] gradient, which is reproduced from Fig.~\ref{fig:gradients}. Each image lists the galaxy's SDSS DR7 object ID, stellar mass, debiased bar vote fraction, and redshift. The SDSS fiber is represented by the black circle at the center of each image with the correct relative scaling; the scale at the bottom right of each image gives the physical size of the 3$\arcsec$ diameter of the SDSS fiber. The images are arranged from lowest redshift (i.e., smallest radius; left) to highest redshift (i.e., highest radius; right). Although the SDSS fiber covers more of the bar structure with redshift, the fiber is still within the extent of the bar at all redshifts. 
\label{fig:gradients_with_stacked_bars}}
\end{figure*}

A similarly important question is ``how much of our bars is covered by the SDSS fiber?'' To illustrate the varying extent of the SDSS fiber with respect to the bar structure as a function of redshift, we present SDSS $r$-band images of typical barred galaxies for each redshift interval at a given stellar mass bin with the SDSS fiber overplotted in Fig.~\ref{fig:gradients_with_stacked_bars}. The physical size of the 3$\arcsec$ SDSS fiber is shown at the bottom right of each image. For each mass bin, we reproduce the [Fe/H] gradients from Fig.~\ref{fig:gradients} for the reader's convenience. 

Clearly, the SDSS fiber covers more of the bar with increasing redshift. This effect is not likely due to an evolution of the bar length: matching our barred galaxies from the gradient analysis to the bar length catalog of \cite{hoyle11}\footnote{$\approx14$\% of our bars in the gradient analysis have bar length measurements from \cite{hoyle11}. This small overlap is because most galaxies in \cite{hoyle11} have significant H$\alpha$ emission, and moreover, they only considered $p_{\rm bar}>0.8$ galaxies.} shows that there is no strong bar length evolution, indicating that the SDSS fiber is truly covering a larger fraction of these bars with redshift. 

To answer the question posed at the beginning of this section more quantitatively, we need bar length measurements for all our barred galaxies. Unfortunately, as was mentioned above, we only have bar lengths for $\approx14$\% of our sample. Fortunately, this subsample is approximately normally distributed across the mass range used in \S\ref{sub:gradient}, thus it is likely to be a representative sample. Assuming that this subsample is indeed representative of our entire bar sample, we find that most of our bars are $3-7$ kpc (in semi-major axis), which is consistent with the findings of other works \citep{erwin05, durbala08, gadotti11}. Thus the SDSS fiber, which ranges from $0.6$ kpc to $2.4$ kpc in radius, covers $\approx8$-$80$\% of the length of the bars in our sample, i.e., the fiber samples a large range of the typical bar lengths in our sample. 

But it is clear that our gradients are limited to well within the bar structure, which is an important caveat to consider and is discussed in more detail in \S\ref{sec:discussion}.

\subsubsection{Stellar population gradients of barred and unbarred galaxies}

Given that our gradients do not extend beyond the bars, here we present the stellar population gradients of barred and unbarred quiescent disk galaxies in Fig.~\ref{fig:gradients}. The top three rows plot [Fe/H], [Mg/Fe], and [N/Fe] as a function of the physical radius of the SDSS fiber (in kpc) in the three bins of stellar mass listed above. The fourth row of Fig. \ref{fig:gradients} superimposes the gradients of all mass bins. At all radii, [Fe/H], [Mg/Fe], and [N/Fe] increases with $\log~M_*$, which is consistent with the bulge trends in Fig.~\ref{fig:stellarpop_vs_mass}.

We stress that the stellar population parameters at a given fiber radius in Fig.~\ref{fig:gradients} are not annular measurements, but rather, integrated measurements, i.e., we model the stellar population of the total area covered by the SDSS fiber, which increases with redshift. Despite the fact that our method of estimating gradients is different from most other works (which use annular measurements), we still find results that are similar to those works. For example, our [Fe/H] gradients are negative (decreases with radius) while our [Mg/Fe] gradients are nearly-flat, both of which are consistent with the findings of past works \citep{mehlert03, sanchez-blazquez07, rawle08, kuntschner10, greene13}. 

We do not consider the stellar age gradients because in addition to probing larger physical distances within galaxies with redshift, we are also probing younger galaxies with redshift. Thus the resulting stellar age gradients can be due to both younger stellar populations at larger physical distances and passive evolution, a degeneracy that we do not attempt to disentangle. 

The main result of this section is evident upon examination of Fig.~\ref{fig:gradients}: there are no differences in the central stellar population gradients of barred and unbarred quiescent disk galaxies that exceed the level of systematic uncertainty. 

\section{Comparison to previous works} \label{sub:comparison}

As summarized in the introduction, previous works in this topic often disagree---some show that bars affect their host galaxies' stellar populations \citep[e.g.,][]{moorthy06, perez11, sanchez-blazquez11, coelho11, williams12} while others show that bars do not \citep[e.g.,][]{cacho14, sanchez-blazquez14}. While some of these differences can be attributed to the choice of stellar population synthesis model, \cite{conroy14} have shown that there is a broad agreement between their model (the one used in this work) and several other popular models, including {\tt EZ\_Ages}  \citep{graves08}, which uses Lick indices as opposed to full-spectrum fitting. Thus assuming that different models produce similar results, we will restrict our comparison to data quantity, data quality, and sample selection.   

Comparing our data quantity to previous works shows that there are only a couple of works that are comparable---\cite{coelho11} have 251 barred galaxies and 324 unbarred galaxies, and \cite{cacho14} have 414 barred galaxies and 1180 unbarred galaxies; both works primarily contained star-forming barred galaxies. Our work contains over a 1,000 quiescent barred galaxies and over 5,000 quiescent unbarred galaxies. 

Comparing our data quality to previous works shows that our stacked spectra have the highest S/N ($\gs100$ \AA$^{-1}$). Previous works have spectra with S/N=10-50 \AA$^{-1}$, which produces less accurate and less precise stellar population parameters. Of course, in order to achieve such high S/N, we have stacked many spectra, which has not been done in any previous work on this topic. Thus our work is measuring the average effects of bars in quiescent disk galaxies. 

Finally, comparing our sample selection to previous works shows that our sample is unique. While we only select galaxies with SDSS spectra that are bereft of two of the strongest optical emission lines---H$\alpha$ and [\ionn{O}{ii}]$\lambda$3727---all previous works selected galaxies with a range of star formation states. In fact, most works require the presence of these lines \citep[e.g.,][]{cacho14}. The main benefit of our selection criteria is that our stacked spectra have minimal emission line contamination, resulting in highly accurate and precise stellar population modeling, which was one of the main goals of this work. This sample selection, however, is an important consideration in our interpretation, which we discuss in \S\ref{sec:discussion} (see also \ref{sub:sample_selection}).  

Based on these three comparisons, it is clear that our work is unlike any of the previous works on this topic. Interestingly though, even with our differences, we find similar results to the latest work on this topic. Namely, like \cite{cacho14}, we find no strong differences in the stellar populations of barred and unbarred galaxies, indicating that bars do not significantly affect the stellar populations of their host galaxies, regardless of their star formation state.

\section{Discussion} \label{sec:discussion}

\subsection{Comparison to predictions of bar-driven secular evolution} \label{sub:secular_evolution}

The results presented in this work indicate that bars do not affect---at least not strongly enough for us to significantly detect---the central stellar populations of quiescent disk galaxies. In the following discussion, we assume that bars are long-lived structures, which is consistent with the latest simulations of bar formation and evolution \citep{athanassoula13a}. Moreover, given that our sample comprises solely of quiescent galaxies, we will restrict our comparison to simulations without star formation. According to these types of simulations from \cite{friedli94}, barred galaxies should have a lower central stellar [Fe/H] (by about $0.20$ dex) compared to unbarred galaxies because the bar-driven inward transportation of stars reduces the initial negative [Fe/H] gradient (see also \citealt{dimatteo13}, who find a reduction of about $0.10$ dex in the central [Fe/H] of barred galaxies, albeit with unrealistic initial conditions). But as shown in Fig.~\ref{fig:stellarpop_vs_mass}, the bulge [Fe/H] of barred galaxies do not differ from that of unbarred galaxies at the few hundredths of a dex level.

Chemical simulations of bar-driven secular evolution without star formation also predict that the slopes of the [Fe/H] gradients of barred galaxies are flatter than those of unbarred galaxies \citep{friedli94, dimatteo13}. But Fig.~\ref{fig:gradients} shows no obvious differences in the [Fe/H] slopes of barred and unbarred galaxies. It is important to note, however, that our gradients do not extend beyond the bar, and are in fact, well within the bars of our sample. This is important because most simulations have shown that the flattening of abundance gradients is most dramatic in the outskirts of galaxies \citep[e.g.,][]{friedli94, dimatteo13}. Nevertheless, simulations predict that the [Fe/H] gradients near the centers are still affected by the presence of a bar---\cite{dimatteo13} predicts a flattening by $\sim$0.01 dex/kpc in the inner regions of their barred galaxy simulation---an effect that we do not observe, albeit this effect may be too subtle for us to detect.  

The above mentioned simulations, however, do not include the effects
of initial star formation and feedback, which are necessary for the accurate
calculation of metallicity gradients. Additionally, quiescent galaxies are not produced
in any simulations, i.e., galaxies are either always star-forming or always quiescent in simulations.
Moreover, since the onset of
bars is around z $\sim 1$ \cite[e.g.,][]{sheth08, kraljic12, melvin14, simmons14},
and assuming that bars are long-lived \citep[e.g.,][]{athanassoula13a}
the simulations should use
initial conditions based on what we
know for galaxies of that era. No observational data are
available for metallicity gradients at such redshifts, apart from
the work of \cite{queyrel12}, which
indicates that some of the gas-phase metallicity gradients of galaxies
around z $\sim 1.2$ could be positive even though most of them are
negative. Furthermore, there are no corresponding measurements
for the stellar metallicity gradients. Finally, these simulations mentioned above assume small bulges, which may optimize the tangential force from bars \citep{laurikainen07}. Thus the predictions from these simulations may be overestimating the influence of bars. In view of
this situation, it does not seem possible to compare adequately theoretical
predictions with our results concerning the differential effect
between barred and unbarred quiescent galaxies at z $\sim0$.

\section{Conclusion} \label{sec:conclusion}

In this work, we use the latest stellar population synthesis model of \cite{conroy12} to estimate the stellar populations of barred and unbarred quiescent disk galaxies with very high S/N ($\gs 100$ \AA$^{-1}$) stacks drawn from the Sloan Digital Sky Survey. Taking advantage of the variable physical size of the SDSS fiber as a function of redshift, we model the stellar populations at different parts within galaxies. We first focus on the bulges of galaxies, which have a typical half-light radius of 1 kpc \citep{fisher10}, and approximates the physical size of the SDSS fiber at $0.02<z<0.04$. We then estimate the stellar population gradients by assuming that galaxies at a fixed stellar mass do not significantly evolve in the redshift range of $0.020<z<0.085$, which spans $\approx 1$ Gyr. Our resulting stellar population gradients extend out to $50\%-70$\% of the median semi-major half-light radius of our samples, which is well within the extent of the bars in our sample.

We find that there are no significant differences in the stellar populations of the bulges or the gradients of barred vs. unbarred quiescent disk galaxies, suggesting that bars are not a significant influence to the chemical evolution of quiescent disk galaxies.

\acknowledgments

This publication has been made possible by the participation of more than 200000 volunteers in the Galaxy Zoo project. Their contributions are individually acknowledged at http://www.galaxyzoo.org/Volunteers.aspx. 

Authors from UC Santa Cruz acknowledge financial support from the NSF Grant AST 08-08133. CC acknowledges funding from NASA grant NNX13AI46G and NSF grant AST-1313280.
EA and AB acknowledge financial support to the DAGAL network from
the People Programme (Marie Curie Actions) of the European
Union's Seventh Framework Programme FP7/2007-2013/ under
REA grant agreement number PITN-GA-2011-289313,
from the CNES (Centre National d'Etudes Spatiales - France)
and from the PNCG (Programme National Cosmologie et Galaxies - France).

Funding for the SDSS and SDSS-II has been provided by the Alfred P. Sloan Foundation, the Participating Institutions, the National Science Foundation, the U.S. Department of Energy, the National Aeronautics and Space Administration, the Japanese Monbukagakusho, the Max Planck Society, and the Higher Education Funding Council for England. The SDSS website is http://www.sdss.org/. The SDSS is managed by the Astrophysical Research Consortium for the Participating Institutions. The Participating Institutions are the American Museum of Natural History, Astrophysical Institute Potsdam, University of Basel, University of Cambridge, Case Western Reserve University, University of Chicago, Drexel University, Fermilab, the Institute for Advanced Study, the Japan Participation Group, Johns Hopkins University, the Joint Institute for Nuclear Astrophysics, the Kavli Institute for Particle Astrophysics and Cosmology, the Korean Scientist Group, the Chinese Academy of Sciences (LAMOST), Los Alamos National Laboratory, the Max-Planck-Institute for Astronomy (MPIA), the Max-Planck-Institute for Astrophysics (MPA), New Mexico State University, Ohio State University, University of Pittsburgh, University of Portsmouth, Princeton University, the United States Naval Observatory, and the University of Washington.

EC deeply thanks Genevieve J. Graves for sharing her stacking code. EC also thanks Jieun Choi and Kevin Bundy for useful discussions. We also thank the anonymous referee for a constructive report. The JavaScript Cosmology Calculator \citep{wright06} was used in the preparation of this paper.

\appendix

\section{Unbarred quiescent disk selection} \label{appen:cheng_selection}

Face-on quiescent disk galaxies, especially those without structures such as bars, are difficult to distinguish from pure elliptical galaxies. In order to illustrate that our unbarred quiescent disk sample is not severely contaminated by pure elliptical galaxies, in this section of the appendix, we compare a sample of spheroid-dominated galaxies to our unbarred quiescent disk sample. 

To select spheroid-dominated galaxies from our quiescent sample (see \S\ref{sub:sample_selection}) we use the criteria outlined by \cite{cheng11}. In their work, they offer a way of selecting early-type, bulge-dominated galaxies without sharp edges, which we consider a satisfactory definition of spheroid-dominated galaxies. The selection method of \cite{cheng11} depends on GIMD bulge+disk decompositions and standard SDSS photometric parameters. In particular, the criteria are:

\begin{enumerate}[noitemsep]

\item $B/T> 0.5$ 
\item $s2> 0.08$
\item $b/a >0.65$
\item $C>2.9$

\end{enumerate}

with $B/T$ representing the bulge-to-total ratio, $s2$ representing the smoothness parameter, $b/a$ representing the $r$ band axis ratio, and $C$ representing the $r$ band concentration. $B/T$ and $s2$ are from the $n=4$ + $n=1$ GIM2D catalog \citep{simard11} while $b/a$ and $C$ are from SDSS DR7. 

\begin{figure*}[t!] 
\centering
\includegraphics[width=\textwidth]{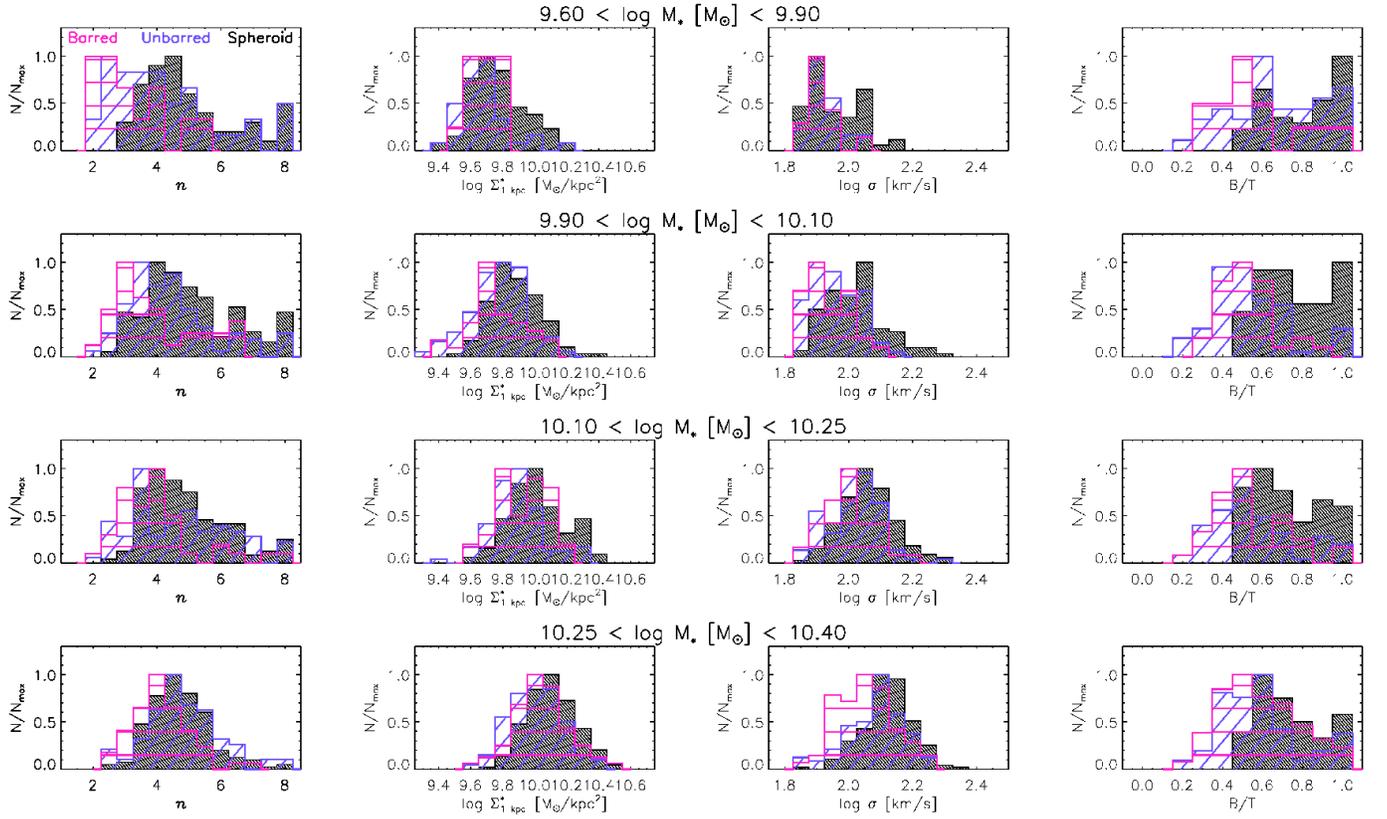}
\caption{Distributions of S\'ersic index ($n$), central surface stellar mass density ($\Sigma_{\rm 1~kpc}^*$), central velocity dispersion ($\sigma$), and bulge-to-total ratio (B/T) for the barred quiescent disks, unbarred quiescent disks, and spheroid-dominated galaxies for the four least massive bins in the bulge analysis (see \S\ref{sub:bulge_pop}). 
\label{fig:faceondisk_vs_elliptical_0}}
\end{figure*}

\begin{figure*}[t!] 
\centering
\includegraphics[width=\textwidth]{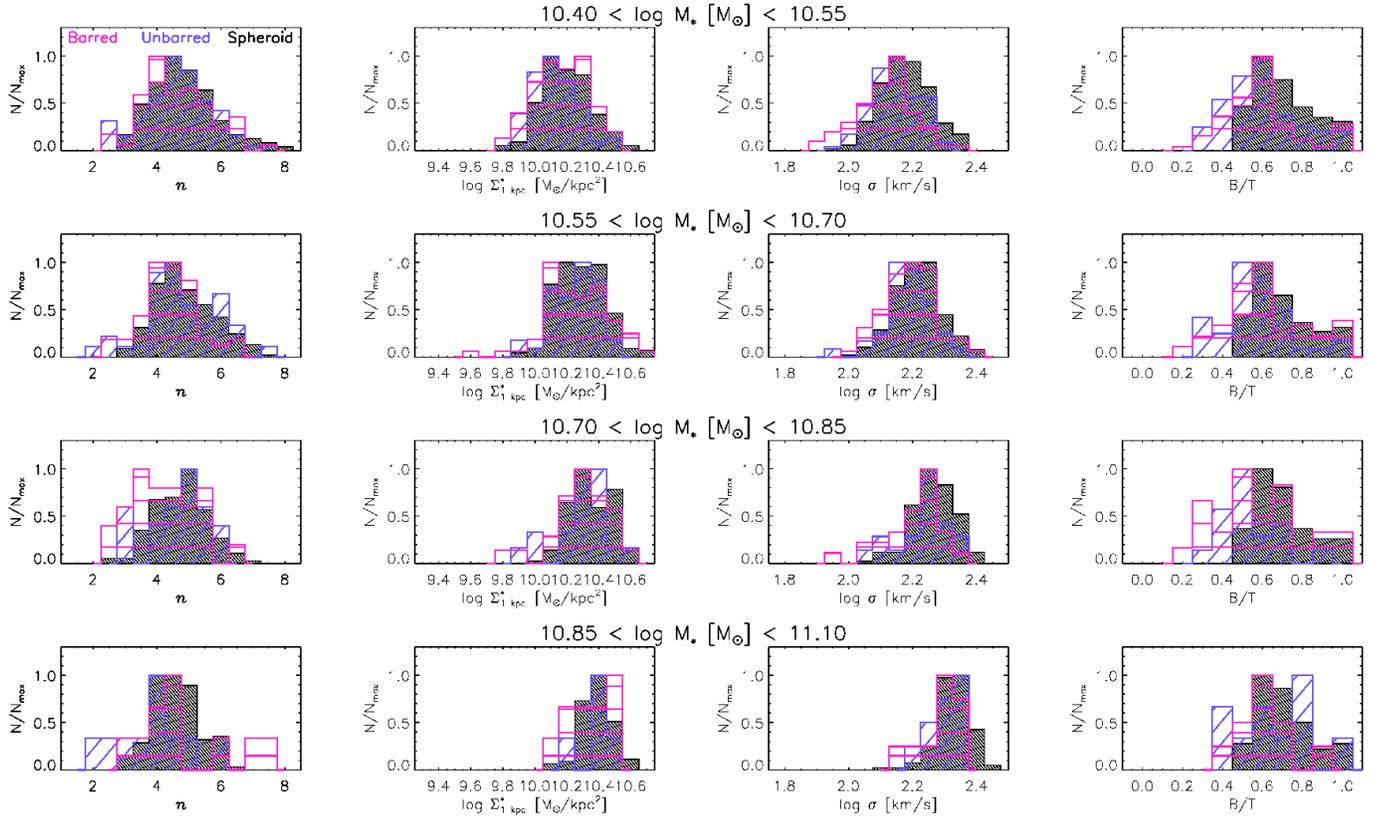}
\caption{Same as Fig.~\ref{fig:faceondisk_vs_elliptical_1} except for the four most massive bins in the bulge analysis.
\label{fig:faceondisk_vs_elliptical_1}}
\end{figure*}

With this spheroid sample, we compare its distributions of global S\'ersic index ($n$), central surface stellar mass density ($\Sigma_{\rm 1~kpc}^*$; \citealt{cheung12}), central velocity dispersion ($\sigma$), and bulge-to-total ratio (B/T) to that of our unbarred quiescent disk sample at the eight mass bins used in our bulge analysis (see \S\ref{sub:bulge_pop}) in Figs.~\ref{fig:faceondisk_vs_elliptical_0} and \ref{fig:faceondisk_vs_elliptical_1}. We also include the distributions of our barred quiescent disk sample for a further comparison. 

These distributions show that the spheroids have, on average, a higher $n$, $\Sigma_{\rm 1~kpc}^*$, $\sigma$, and B/T than both the barred and unbarred quiescent disks for the four least massive bins, indicating that our unbarred quiescent disks are not dominated by spheroid-dominated galaxies.  

For the more massive bins, however, their distributions become more similar, indicating that the structural properties of these classes of galaxies are almost indistinguishable at high masses. This similarity though, does not mean that our unbarred quiescent disk sample is severely contained by ellipticals. Indeed, the structural distributions of our barred quiescent disk sample are also similar to that of the spheroid-dominated sample at these massive bins. And since bars are pure disk phenomena \citep{athanassoula13b}, we conclude that the structural parameters we considered simply cannot distinguish spheroid-dominated galaxies from quiescent disk galaxies at high stellar masses. Therefore, given the ineffectiveness of these structural parameters to distinguish face-on quiescent disks from spheroid-dominated galaxies at high masses, we rely on the visual morphologies of Galaxy Zoo.

Perhaps the most reassuring result of this section, however, is that the structural distributions of our barred and unbarred quiescent face-on disks are similar, indicating that we are selecting similar types of galaxies.

\section{Comparison of Barred vs. Unbarred Stacks}  \label{appen:stacks}
\begin{figure*}[t!] 
\centering
\includegraphics[scale=.6]{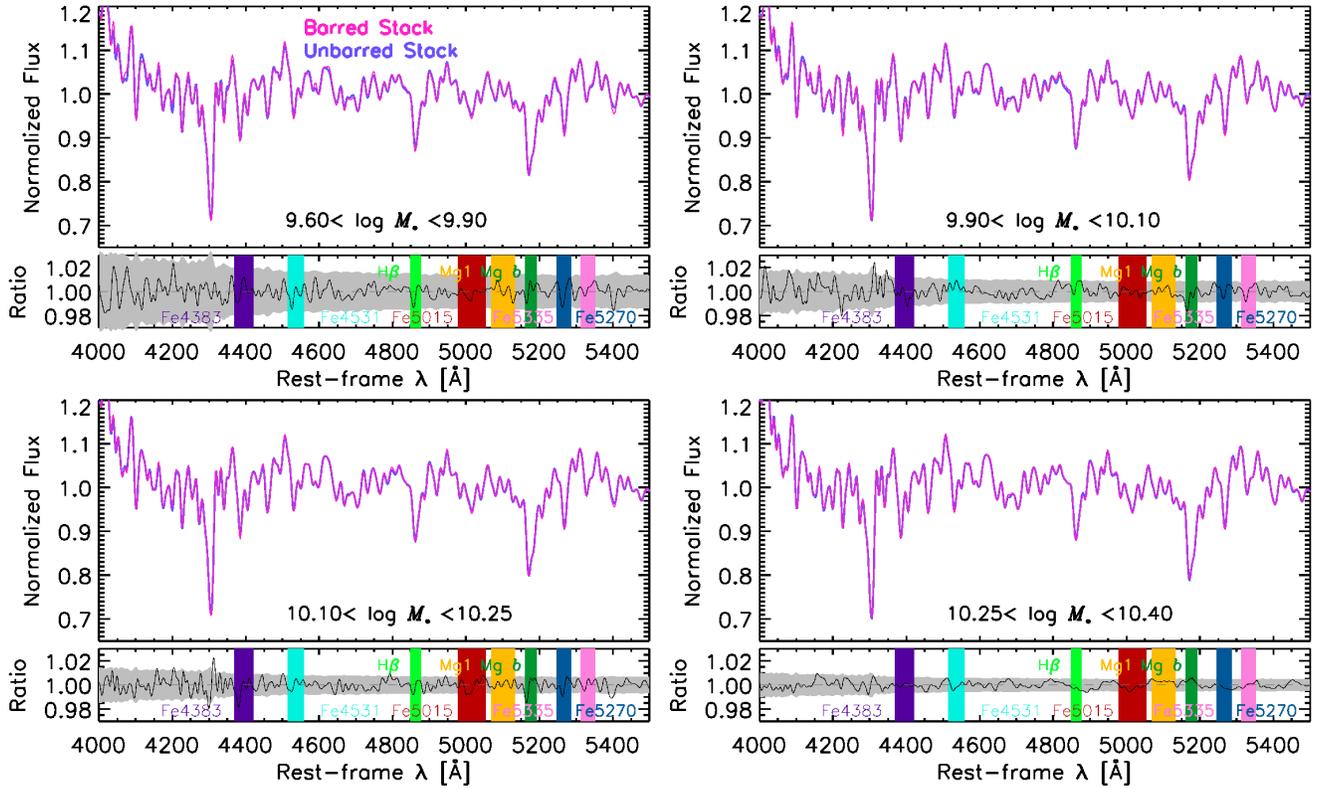}
\caption{Comparison of barred and unbarred stacked spectra for the four least massive bins in the bulge analysis (see \S\ref{sub:bulge_pop}). The ratio of the barred to unbarred stacks are shown below the stacked spectra, with the grey region representing its error; several Lick indices highlighted. 
\label{fig:stack_comparison_0}}
\end{figure*}

\begin{figure*}[t!] 
\centering
\includegraphics[scale=.6]{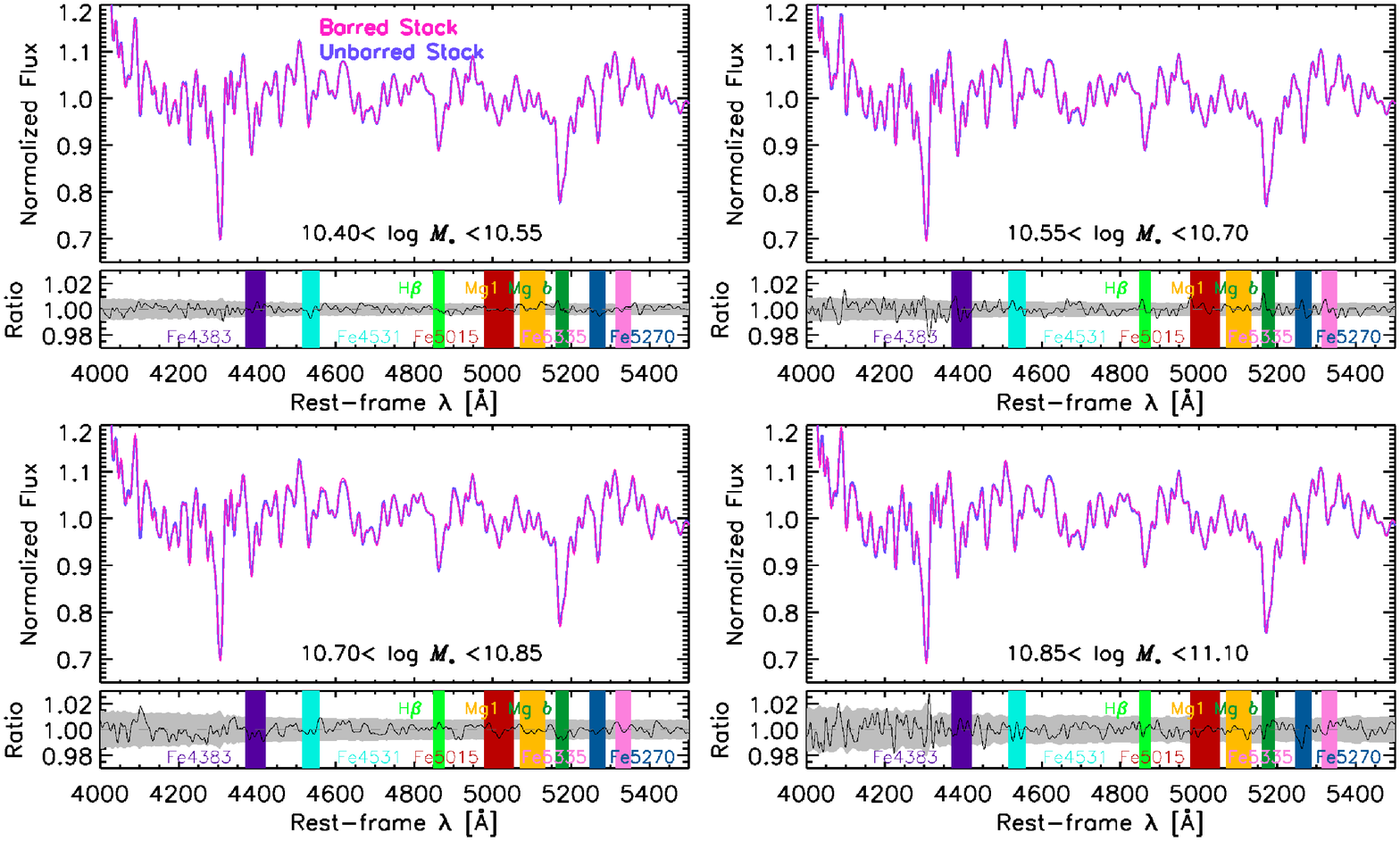}
\caption{Same as Fig.~\ref{fig:stack_comparison_0} but for the four most massive bins in the bulge analysis.
\label{fig:stack_comparison_1}}
\end{figure*}

In this section we directly compare the stacked spectra of the barred and unbarred samples. Figs.~\ref{fig:stack_comparison_0} and \ref{fig:stack_comparison_1} plot the barred stacks over the unbarred stacks for each stellar mass bin analyzed in the bulge analysis (\S\ref{sub:bulge_pop}). The bottom panels show the ratio of the barred to unbarred stacks, highlighting some relevant Lick indices. 

\begin{figure*}[t!] 
\centering
\includegraphics[scale=.6]{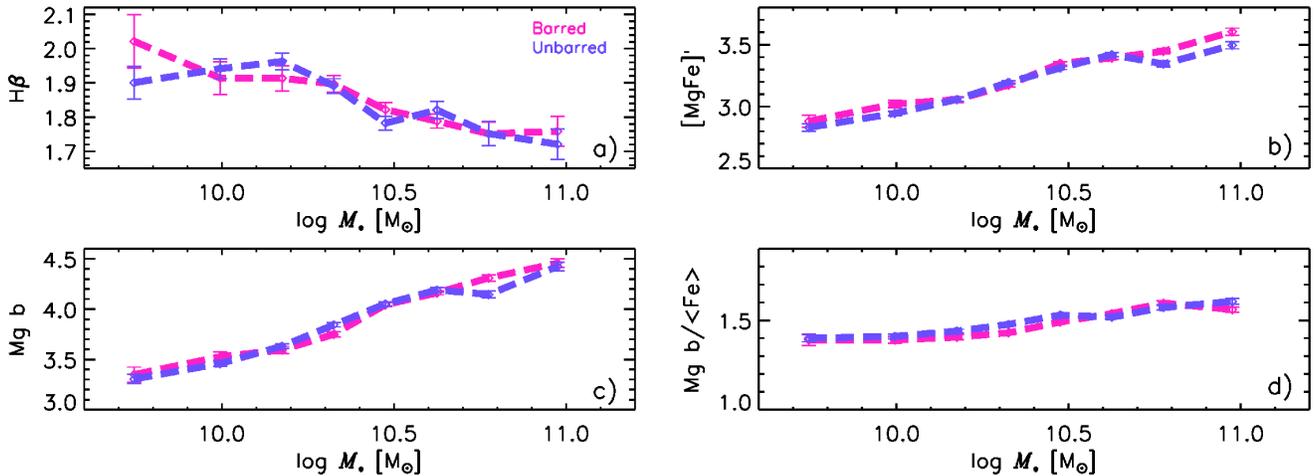}
\caption{Lick indices vs. stellar mass of the bulge analysis (\S\ref{sub:bulge_pop}). Namely, {\it a):} H$\beta$ vs. $\log~M_*$, {\it b):} [MgFe]' vs. $\log~M_*$, {\it c):} Mg b vs. $\log~M_*$, and {\it d):} Mg b/<Fe> vs. $\log~M_*$ at $0.02<z<0.04$. 
\label{fig:lick_indices}}
\end{figure*}

Clearly, there are very small differences in these stacks. To summarize these differences, we plot several Lick indices vs. stellar mass in Fig.~\ref{fig:lick_indices}. H$\beta$, a tracer of the stellar age, [MgFe]', a tracer for metallicity \citep{thomas03}, and Mg b and Mg b/<Fe>, a tracer for the $\alpha$-abundance \citep{thomas03}, all show no significant differences between barred and unbarred galaxies, mirroring our main result in \S\ref{sub:bulge_pop}. 
 
\section{Comparison of Model Fits to Input Spectra}  \label{appen:residuals}

\begin{figure*}[t!] 
\centering
\includegraphics[scale=.6]{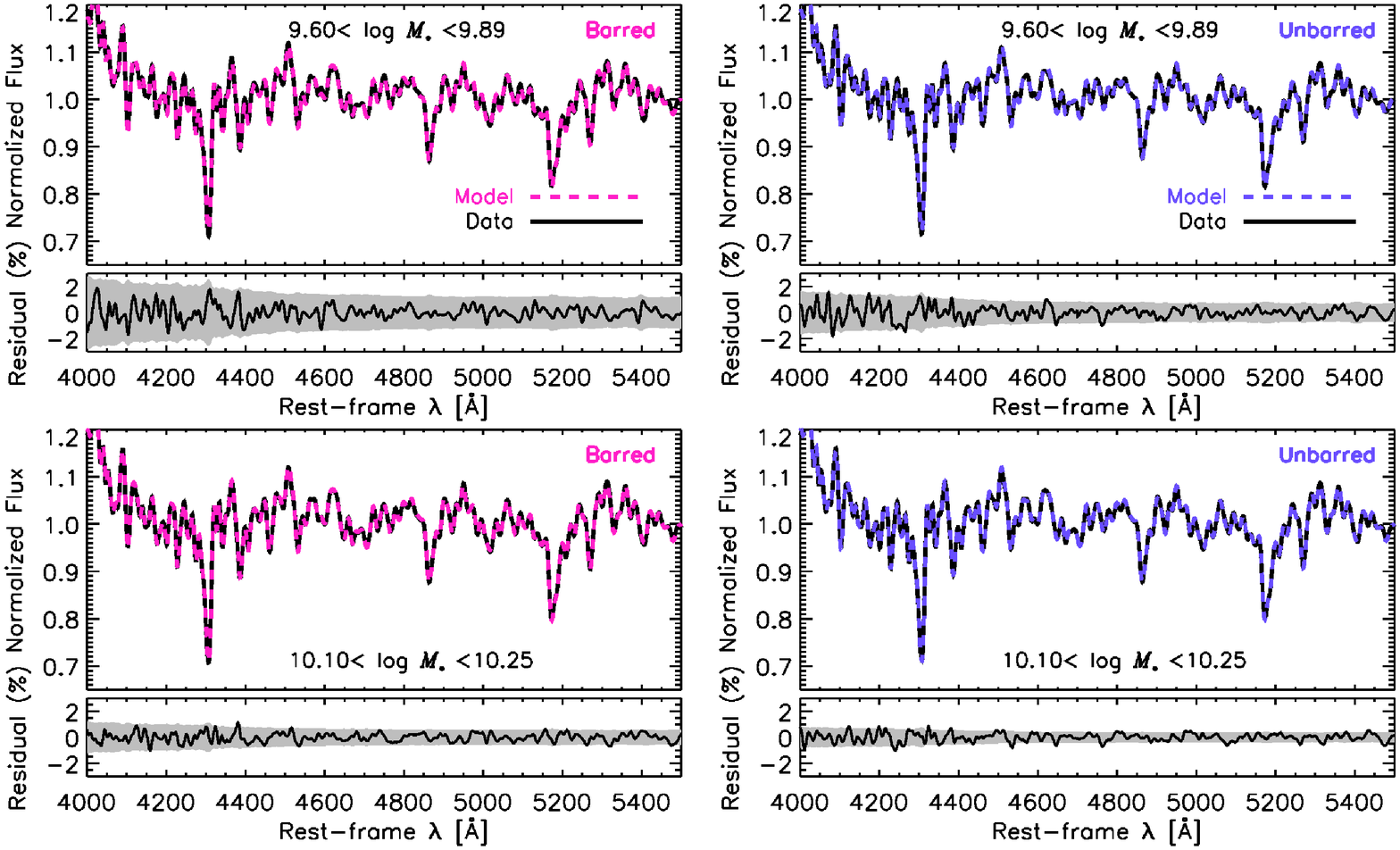}
\caption{Comparison of model spectra and input spectra for two mass bins from the bulge analysis (\S\ref{sub:bulge_pop}). Barred and unbarred stacks are on the left and right, respectively. The residual of the model and the input stacked spectra are displayed below each spectra, with the grey region representing the noise of the stacked spectra. The residuals are small and within the noise, indicating that the fits are excellent. 
\label{fig:spectra_comparison_0}}
\end{figure*}

\begin{figure*}[t!] 
\centering
\includegraphics[scale=.6]{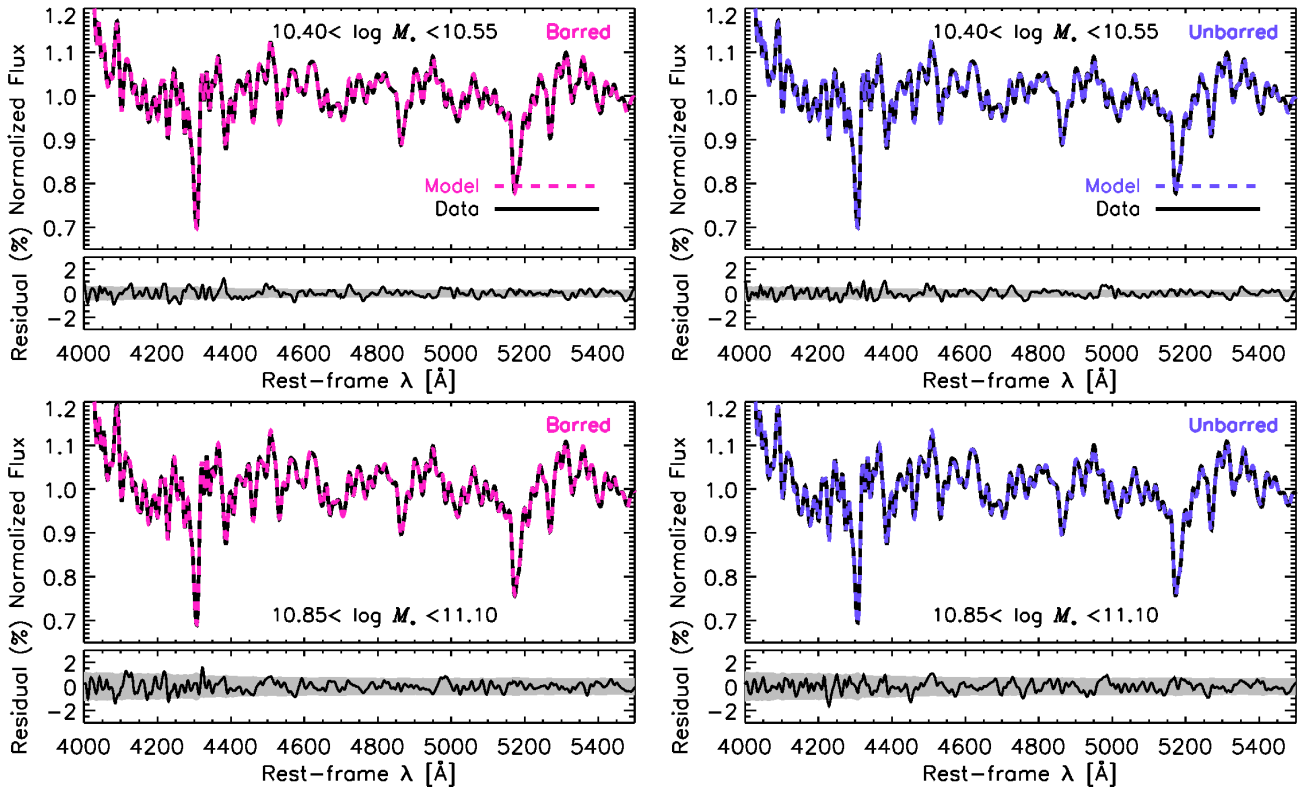}
\caption{Same as Fig.~\ref{fig:spectra_comparison_1} except for two more massive bins. 
\label{fig:spectra_comparison_1}}
\end{figure*}

In this section of the appendix, we illustrate the quality of the model fits by showing the input stacked spectra and the model spectra in Figs.~\ref{fig:spectra_comparison_0} and \ref{fig:spectra_comparison_1} for four mass bins from the bulge analysis (see \S\ref{sub:bulge_pop}); the barred stacks are on the left and unbarred stacks are on the right. At the bottom of each spectra, we show the residuals of the model and input spectra. The first feature to note is that the residuals are small, about one percent, indicating that the fits are excellent. Secondly, the residuals for the barred and unbarred model spectra are similar, indicating that the quality of the fits of the two samples are comparable.

\end{document}